\documentclass[a4paper,12pt]{article}
\usepackage{lmodern}
\usepackage{amssymb,amsmath,amsthm}
\usepackage{ifxetex,ifluatex}
\usepackage{graphicx}
\usepackage[round]{natbib}
\usepackage{here}
\usepackage{color}
\usepackage{comment}
\usepackage{authblk}
\usepackage{hyperref}

\title{Maximum Likelihood Estimation of the Vector AutoRegressive To Anything (VARTA) model}
\author[1]{Jonas Andersson}
\author[2]{Dimitris Karlis}

\affil[1]{Department of Business and Management Science, Norwegian School of Economics}
\affil[2]{Department of Statistics, Athens University of Economics and Business}

\begin{document}

\maketitle

\newtheorem{example}{Example}

\begin{abstract}

The literature on multivariate time series is, largely, limited to either
models based on the multivariate Gaussian distribution or models specifically developed for a given application. In this paper we develop a general approach which is based on an underlying, unobserved, Gaussian Vector Autoregressive (VAR) model. Using a transformation, we can capture the time dynamics as well as the distributional properties of a multivariate time series. The model is called the Vector AutoRegressive To Anyting (VARTA) model and was originally presented by \cite{biller2003modeling} who used it for the purpose of simulation. 
In this paper we derive a maximum likelihood estimator for the model and investigate its performance. We also provide diagnostic analysis and how to compute the predictive distribution. The proposed approach can provide better estimates about the forecasting distributions which can be of every kind not necessarily Gaussian distributions as for the standard VAR models.

\end{abstract}

\noindent {\textbf{Keywords}: {non-Gaussian time series; maximum likelihood estimation; predictive distribution}

\vspace{3mm}

\noindent {\textbf{JEL classification}: {C13; C22; C58}

\section{Introduction}

There is an enormous literature on how to model and forecast multivariate time series.
 Many multivariate statistical models are based on the, analytically tractable, multivariate Gaussian distribution.
On the other hand,  there is a much smaller literature for multivariate time series
for non-Gaussian data, including the problem of forecasting such data. One reason might be the substantial difficulty to define such models
and examine their properties. The present paper aims at filling  this gap by considering multivariate time series models for 
non-Gaussian data. 

The derivation of the model used in this paper is based on an extension of the NORTA
(normal to anything) approach, see \cite{cario1997modeling}. Starting from a Gaussian Vector Autoregressive  (VAR) model,
we apply the probability integral transformation (PIT) to the data
to end up with correlated multivariate uniform random variables.
This transformation keeps both the time and cross correlations.
 Then, by the inverse transformation of the cumulative distribution function (CDF) we transform the data to any distribution, still keeping the
structure of both the time and cross correlations.  Since we can
apply a different transformation to each marginal, such an approach
can lead to a variety of different marginal models keeping the
autoregressive structure and the correlation between the series. Such an
approach allows to consider very flexible families of continuous
time series models transferred to the case of any type of  time
series.

The model is called the Vector AutoRegressive To Anyting (VARTA) model and was originally presented by \cite{biller2003modeling} who used it for the purpose of simulation. See also \cite{biller2009copula} for further discussion on this approach. 
\cite{zhang2022margin} examined conditions that a Gaussian 
VAR($k$) model  has univariate margins that are
autoregressive of order $k$ or lower-dimensional margins that are also VAR($k$). Such conditions can help to built the model by selecting appropriate univariate marginal distributions. 

VARTA can be very helpful to formulate time series models with non-Gaussian marginals. The existing models in the literature are either very specialized, based on a particular multivariate distribution, or unnecessarily flexible, like copula approaches can sometimes become. The VARTA approach falls in between in these extremes in the sense that it models time dynamics as the familiar VAR model while allowing any choice of continuous marginal distributions.

For continuous non-Gaussian models, there have been proposed models for bivariate
exponential \citep{dewald1989bivariate,block1988bivariate}, bivariate uniform
\citep{ristic2003bivariate},  bivariate beta and gamma
\citep{bakouch2009bivariate}, bivariate Weibull of the
Marshal-Olkin type \citep{jose2011marshall}, bivariate semi
Pareto (\cite{Thomas2004}) and multivariate Weibull
\cite{yeh2013general}. Finally \cite{CJS:CJS385} proposed
state-space models for multivariate longitudinal data of mixed
types. It is  important to note that in all cases the marginal
models are of the same kind. \cite{fokianos2021multivariate} and  \cite{livsey2018multivariate}
provide interesting contributions for the case of multivariate count time series while an example of a bivariate time series model in $\mathbb Z$, rather than $\mathbb Z_+$, is given in 
\cite{Bullaetal2017}.


Forecasting of transformed time series was also treated quite early in the literature for the univariate case by
\cite{granger1976forecasting}, They considered the problem where a forecasting model is available for the Gaussian process $X_t$ 
which is stationary (or can be made so after differencing) 
but their interest centers on the instantaneous transformation $Y_t = T(X_t)$. They
examined the autocovariance structure of and methods for forecasting the transformed series
$Y_t= T(X_t)$ for a transformation $T(\cdot)$. 
In this paper we consider a vector autoregressive (VAR) model  which is, subsequentally, transformed with a particular transformation based on inverse cumulative distribution functions (CDFs).

The idea is that starting from a well known model like the
VAR we end up with time series which are correlated and keep the
autoregressive structure. 
Such a transformation is not new in the literature, in the sense
that it has been used by many other authors.  For example
\cite{cario1996autoregressive} called such an approach for continuous time series autoregressive to anything (ARTA). \cite{erhardt2010sampling} used the approach for sampling discrete
values with given correlation. The key idea is a univariate in
such cases transformation from  Gaussian distribution to any
other. \cite{masarotto2012gaussian} described a general class of
models based on this transformation that contains a multitude of
models. See also the recent work of \cite{duker2024high} and  \cite{kim2025latent} on such latent Gaussian models.

The contribution of the present paper is that we present a full parametric inference approach for VARTA models. 
An advantage is that it uses the huge
theory about likelihood inference on VAR processes for random vectors with arbitrary
marginal distribution, even mixtures of more than one. Note that
the marginal $F_j(\cdot)$ does not even need to be a known
distribution: it can be directly estimated (parametricaly or not)
by the data, as long as the estimated CDF can be inverted.
Furthermore, we can model a wide range of time series.  One more
advantage is that we can simulate very easily by following the
inverse process while the forecasting distribution can be easily set via simulation or even closed forms
based on the assumed transformation. 

We demonstrate the idea with data from wind speed assuming Weibull marginal distributions.
The properties of such an approach are discussed.
The remainder of the paper is as follows: 
Section \ref{the-model} defines the VARTA model while Section \ref{inference} describes ML estimation, asymptotic properties of the estimator, diagnostics and forecasting. Section \ref{the-2-dimensional-case} presents an illustration with Weibull marginal distributions using wind data. Section \ref{simul} provides
simulation evidence for the finite sample properties of the estimator and its asymptotic approximation of standard error, including higher dimensional data. Concluding remarks can be found in Section \ref{conclude}.

\section{The model}\label{the-model}

Consider the covariance stationary \(p\)-dimensional vector
autoregressive (VAR) model or order one, denoted as VAR(1), given by
\begin{equation}
\boldsymbol{Z}_t=A\boldsymbol{Z}_{t-1}+\boldsymbol{\eta}_t,
\label{eq:var1}
\end{equation}
 For identification purposes, we
insist that the unconditional variances of all components of
$\boldsymbol{z}_t$ are unity, then the restrictions on $\Omega$, namely that
\begin{equation}
\Omega=\Sigma-A\Sigma A',
\label{eq:unity}
\end{equation} follow. 

Here, \begin{equation}
\Sigma=\left(
\begin{aligned}
1 && \rho_{12} && \cdots && \rho_{1p}\\
\rho_{12} && 1 && \cdots && \rho_{2p}\\
\vdots && \vdots && \ddots && \vdots\\
\rho_{1p} && \rho_{2p} && \cdots && 1\\
\end{aligned}
\right).
\end{equation}

Going back to the \(p\)-dimensional VAR(1)-model we transform all or
some of the components so that they are allowed to follow a
pre-specified marginal distribution. This is done by first applying the
univariate Gaussian CDF, $\Phi(\cdot)$, to
the components of $\boldsymbol{Z}_t$, namely
\begin{equation}
U_{it}=\Phi(Z_{it}),
\label{eq:utrans}
\end{equation} for $i=1,\ldots,p$ and $t=1,\ldots,n$. These are now
following a uniform $U(0,1)$ distribution and can, in turn, be
transformed to any continuous, distribution $F_i$, by using the
quantile function (the inverse of the CDF, $F_i^{-1}$).
\begin{equation}
X_{it}=F_i^{-1}(u_{it})=F_i^{-1}(\Phi(Z_{it}))
\label{eq:Ftrans}
\end{equation} for $i=1,\ldots,p$ and $t=1,\ldots,n$.

The $p$-dimensional stochastic variables $\boldsymbol{X}_t$ are the
stochastic variables, which we believe our observations are realizations
of. They have temporal dynamics determined by the underlying VAR(1)-model
and the components have the marginal distributions $F_i$,
$i=1,\ldots,p$.

\section{Inference}
\label{inference}
\subsection{Estimation of the model}\label{estimation-of-the-model}

A full maximum likelihood estimation can be done in the following way.
By first writing \begin{equation}
\boldsymbol{X}=\boldsymbol{g}(\boldsymbol{Z})
\end{equation} where the boldfaced letters indicate a transformation of
a multidimensional function of a multidimensional variable. Written out,
we have the \(np \times 1\)-vector \begin{equation}
\boldsymbol{g}(\boldsymbol{Z})=
\left(
\begin{gathered}
g_1(Z_{11})\\
\vdots\\
g_1(Z_{1n})\\
\vdots\\
g_p(Z_{p1})\\
\vdots\\
g_p(Z_{pn})\\
\end{gathered}
\right),
\end{equation} and then define \(\boldsymbol{g}\) by \begin{equation}
g_i(z)=F_i^{-1}(\Phi(z)).
\label{eq:inverse}
\end{equation}
We note that the inverse of \(\boldsymbol{g}\) is
\begin{equation}
\boldsymbol{g}^{-1}(\boldsymbol{X})=
\left(
\begin{gathered}
\Phi^{-1}(F_1(X_{11}))\\
\vdots \\
\Phi^{-1}(F_1(X_{1n}))\\
\vdots\\
\Phi^{-1}(F_p(X_{p1}))\\
\vdots\\
\Phi^{-1}(F_p(X_{pn}))\\
\end{gathered}
\right).
\end{equation} Since each component in \(\boldsymbol{g}\) is a monotone
transformation we can derive that the probability density of
\(\boldsymbol{X}\), say \(f_{\boldsymbol{X}}\), can be written
\begin{equation}
f_{\boldsymbol{X}}(\boldsymbol{x})=
f_{\boldsymbol{Z}}(\boldsymbol{g}^{-1}(\boldsymbol{x}))|J|,
\end{equation} where \(f_{\boldsymbol{Z}}\) is the probability density
of the \(np\)-dimensional Gaussian distribution with mean
\(\boldsymbol{0}\) and block-diagonal \(np \times np\)
variance-covariance matrix, \(\textnormal{diag}(\Sigma)\) and \(|J|\) is
the determinant of the Jacobian of
\(\boldsymbol{g}^{-1}(\boldsymbol{x})\).

Since each element in \(\boldsymbol{g}^{-1}(\boldsymbol{x})\) is a
function of one variable only, the Jacobian becomes diagonal and thus
\begin{equation}
|J|=\prod_{i=1}^p
\prod_{t=1}^n \frac{\partial \Phi^{-1}(F_i(x_{it}))}{\partial x_{it}},
\label{eq:jacobian}
\end{equation} which seldom has an analytic expressions but which is
feasible to compute numerically. The log likelihood \begin{equation}
\log L(A,\boldsymbol{\rho},\boldsymbol{\theta})=\log f_{\boldsymbol{Z}}(\boldsymbol{g}^{-1}(\boldsymbol{x}))+\sum_{i=1}^p \sum_{t=1}^n \log  \frac{\partial \Phi^{-1}(F_i(x_{it}))}{\partial x_{it}},
\label{eq:loglik}
\end{equation} where \(\boldsymbol{\rho}\) is a \(p(p-1)/2\)-vector
containing the unconditional correlations between the components in
\(\boldsymbol{Z}_t\), and \(\boldsymbol{\theta}\) are the parameters in
the marginal distributions \(F_i\), \(i=1,...,p\).

For the VAR(1) case we can note that
\(f_{\boldsymbol{Z}}(\boldsymbol{z})\) can be computed by
\begin{equation}
f_{\boldsymbol{Z}}(\boldsymbol{z})=f_{\boldsymbol{Z}_1}(\boldsymbol{z}_1)\prod_{t=2}^n f_{\boldsymbol{Z}_t|\boldsymbol{Z}_{t-1}}(\boldsymbol{z}_t|\boldsymbol{z}_{t-1}),
\end{equation} where
\(\boldsymbol{Z}_1 \sim N_p(\boldsymbol{0},\Sigma)\) and
$\boldsymbol{Z}_t|\boldsymbol{Z}_{t-1}=\boldsymbol{z}_{t-1} \sim N_p(A\boldsymbol{z}_{t-1},\Omega)$.

\subsection{The VAR(k)-case}
\label{VARk}
The extension to more than one lag is only marginally more complicated than the lag one case. The transformation from the Gaussian model \eqref{eq:var1} to the non-Gaussian marginal models, \eqref{eq:utrans}-\eqref{eq:Ftrans}, are the same for the multi-lag case.  The only complication is that that formula \eqref{eq:unity} has to be augmented in order to ensure that the marginal variances of $\boldsymbol{Z}_t$ are all one. For a VAR(k)-model we make the unconditional variances unity in the following way. First, we write the $p$-dimensional VAR(k)-model as a $kp$-dimensional VAR(1)-model

\begin{equation}
\left(
\begin{aligned}
& \boldsymbol{Z}_t\\
& \boldsymbol{Z}_{t-1}\\
& \vdots\\
& \boldsymbol{Z}_{t-k+1}
\end{aligned}
\right)
=
\left(
\begin{aligned}
& A_1 && A_2 && \cdots && A_{k-1} && A_k\\
& I_p && 0 && \cdots && 0 && 0\\
& 0 && I_p && \cdots && 0 && 0\\
& \vdots && \vdots && \ddots & \vdots && \vdots\\
& 0 && 0 && \cdots && I_p && 0
\end{aligned}
\right)
\left(
\begin{aligned}
& \boldsymbol{Z}_{t-1}\\
& \boldsymbol{Z}_{t-2}\\
& \vdots\\
& \boldsymbol{Z}_{t-k}
\end{aligned}
\right)
+
\left(
\begin{aligned}
& \boldsymbol{\eta}_t\\
& 0\\
& \vdots\\
& 0
\end{aligned}
\right)
\label{eq:companion}
\end{equation}
We represent \eqref{eq:companion} with the more compact notation
\begin{equation}
\boldsymbol{W}_t = B \boldsymbol{W}_{t-1} + \boldsymbol{\varepsilon}_t
\end{equation}
and note that the variance-covariance matrix of $\boldsymbol{W}_t$ is
\begin{equation}
\Sigma_k = \left(
    \begin{aligned}
        & \Gamma_0 && \Gamma_1 && \cdots && \Gamma_{k-1}\\
        & \Gamma_1' && \Gamma_0 && \cdots && \Gamma_{k-2}\\
        & \vdots && \vdots && \ddots & \vdots\\
        & \Gamma_{k-1}' && \Gamma_{k-2}' && \cdots && \Gamma_0\\
    \end{aligned}
    \right)
\end{equation}
where $\Gamma_s=C(\boldsymbol{Z}_t,\boldsymbol{Z}_{t-s})$ and all elements on the diagonal of $\Gamma_0$ are equal to one.
Now, analogously to \eqref{eq:var1}, we can ensure that all unconditional variances of $\boldsymbol{Z}_t$ is unity by computing
\begin{equation}
\Theta = \Sigma_k-B\Sigma_k B'
\end{equation}
and extracting the upper left $p \times p$-matrix to obtain $\Omega$. $\Omega$ can then be seen to be 
\begin{equation}
\Omega = \Gamma_0 - \left( \sum_{i=1}^k A_i \Gamma_{i-1}'A_1' + \sum_{r=2}^{k-1} \left(\sum_{i=1}^{r-1} A_i \Gamma_{r-1} + \sum_{i=r}^{k-1} A_i \Gamma_{i-r}'\right)'A_r'+\sum_{i=1}^k A_i \Gamma_{k-i}A_k'\right)
\end{equation}
A conditional maximum likelihood estimator can now be found by maximizing
\begin{equation}
\log L(A_1,\ldots,A_k,\boldsymbol{\rho},\boldsymbol{\theta})=\sum_{t=k+1}^n \log h(\boldsymbol{g}^{-1}(\boldsymbol{x}_t))+\sum_{i=1}^p \sum_{t=k+1}^n \log  \frac{\partial \Phi^{-1}(F_i(x_{it}))}{\partial x_{it}},
\label{eq:loglikmulti}
\end{equation}
where $h$ is the conditional density function for $\boldsymbol{Z}_t$ given $\boldsymbol{Z}_{t-1} =\boldsymbol{z}_{t-1},...,\boldsymbol{Z}_{t-k}=\boldsymbol{z}_{t-k}$, i.e. a $N_p(A_1\boldsymbol{z}_{t-1}+...+A_k \boldsymbol{z}_{t-k},\Omega)$ distribution. 

\subsection{Asymptotic normality}
\label{sec:asymp}

Conditions for asymptotic normality of the maximum likelihood estimator can be found by following the theory of \cite{tjostheim1986estimation} and \cite{ling2010general}. We state the conditions here below.

Consider the log-likelihood defined in \eqref{eq:loglik}. One can see that it can be factorized into two different parts. The first is a multivariate Gaussian log-likelihood applied to the inverse transformation $\boldsymbol{g}^{-1}$ of the observations, and the second the sum of the derivatives of the inverse transformations, $g_i$, $i=1,\ldots,p$. In order to show the consistency and the asymptotic normality of the ML estimates some  regularity restrictions must be fulfilled. In order to represent all parameters in one single vector we introduce $\boldsymbol{\beta}=(A,\boldsymbol{\rho},\boldsymbol{\theta})'$.

In the case of stationary, ergodic and marginally normally distributed observations, the estimator is clearly consistent and asymptotically Gaussian, following, e.g., from \cite{tjostheim1986estimation} or \cite{ling2010general}. For this case, the second part of the log-likelihood in \eqref{eq:loglik} becomes zero. The additional assumptions that has to be made for other cases are therefore solely on the properties of the marginal CDFs $F_i, i=1,\ldots,p$. 

\vspace{0.5cm}

\noindent {\bf Theorem:} If conditions (i)-(v) below are satisfied then the MLE are consistent and asymptotically they follow a Gaussian distribution. The conditions are:

\begin{itemize}
    \item[
(i)] the true parameter $\boldsymbol{\beta}_0$ lies in the interior of the parameter space and the model is identifiable;  
\item[(ii)] the marginal distribution functions $F_i$, $i=1,\ldots,p$, are continuous, strictly increasing on their support, and sufficiently smooth so that their densities $f_i=F_i'$ exist, are bounded away from zero and infinity on compact subsets of the support, and allow differentiation under the integral sign; 
\item[(iii)] the conditional log-likelihood is twice continuously differentiable in a neighborhood of 
$\boldsymbol{\beta}_0$ with suitable bounds on its derivatives so that dominated convergence and uniform laws of large numbers apply and 
\item[(iv)] the marginal distributions should be such that the process is weakly dependent in the sense of $\alpha$-mixing with mixing coefficients decaying sufficiently fast so that a a central limit theorem apply for the score; and 
\item[(v)] the Fisher information matrix at $\boldsymbol{\beta}_0$ is finite and nonsingular. 
\end{itemize}

\noindent {\it Proof:  }
Under the conditions (i)-(v), a Taylor expansion of the first-order conditions can be derived. This together with the CLT for the score and Slutsky’s theorem gives that $\sqrt{n}(\boldsymbol{\hat{\beta}}-\boldsymbol{\beta}_0)$ is asymptotically Gaussian with covariance matrix given by the inverse of the Fisher information. These conditions are in line with the framework of \cite{tjostheim1986estimation} for nonlinear time series maximum likelihood estimation. 

\vspace{0.5cm}

We would like to point that the conditions above primarily prevent from singular cases. For example, if the parameter is on the boundary of the parametric space the distribution may collapse to a one-point distribution which can create problems with the covariance matrices.
Also cases when any dimension is collapsing are excluded. The differentiability of the CDF is necessary for being able to derive the Fisher information matrix

In Section \ref{simul} we investigate these asymptotic properties for finite samples by means of a Monte Carlo study.

\subsection{Residual analysis} \label{sec:resid}

Testing of model assumptions can be done in a relatively straight-forward fashion in the following way. The transformations $\hat{Z}_{it}=\Phi^{-1}(\hat{F}_i(X_{it}))$, $i=1,\ldots,p$, are used to compute residuals
\begin{equation}
    \hat{\boldsymbol{\eta}}_{t}=\hat{\boldsymbol{Z}}_t-A\hat{\boldsymbol{Z}}_{t-1}
\end{equation}
which can be used to test the temporal assumptions of the model. If the model is correctly specified the residuals should lack auto- and cross-correlations. Furthermore, if the distributional assumptions of the model are correct $\hat{\boldsymbol{Z}}_t$ should follow a multivariate Gaussian distribution with expectations zero and variances one. For small samples, adjustments might have to be made to account for estimation uncertainty since the assumptions are made on $\boldsymbol{\eta}_t$ and $\boldsymbol{Z}_t$, not $\hat{\boldsymbol{\eta}}_{t}$ and $\hat{\boldsymbol{Z}}_t$.

\subsection{Forecasting}
\label{forecasting}

Assume that we have a sample $\boldsymbol{X}_1,\boldsymbol{X}_2,...,\boldsymbol{X}_n$.  
Forecasting, using the estimated model,  is done by first forecasting $\boldsymbol{Z}_{n+h}$ and then transforming to a forecast for $\boldsymbol{X}_{n+h}$. The forecast distribution is found using simulation. Let us call the potential future trajectories of $\boldsymbol{Z}_t$ and $\boldsymbol{X}_t$, $\hat{\boldsymbol{Z}}_t$ and $\hat{\boldsymbol{X}}_t$, respectively. Firstly $M$  one-step-ahead forecasts are simulated by 
\begin{equation}
    \begin{aligned}
    & \hat{\boldsymbol{Z}}^{(j)}_{n+1}=\hat{A}\hat{\boldsymbol{Z}}_{n}+\hat{\boldsymbol{\eta}}^{(j)}_{n+1}\\
    & \hat{X}^{(j)}_{i,n+1}=F_i^{-1}(\Phi(\hat{Z}_{i,n+1})), i=1,\ldots,p
    \end{aligned}
    \label{eq:forecastdist}
\end{equation}
   for $j=1,\ldots,M$. 
   This is then a sample from the one-step-ahead forecast distribution. To get a sample from the h-step-ahead forecast distribution, we apply a slightly modified version of equation (\ref{eq:forecastdist})
\begin{equation}
    \begin{aligned}
    & \hat{\boldsymbol{Z}}^{(j)}_{n+k}=\hat{A}\hat{\boldsymbol{Z}}^{(j)}_{n+k-1}+\hat{\boldsymbol{\eta}}^{(j)}_{n+k}\\
    & \hat{X}^{(j)}_{i,n+k}=F_i^{-1}(\Phi(\hat{Z}_{i,n+k})), i=1,\ldots,p
    \end{aligned}
\label{eq:forecastdist2}
\end{equation}
for $j=1,\ldots,M$ and  $k=2,\ldots,h$.



One way to find forecasts of $Y$, is to apply the inverse of the transformation $T(\cdot)$  to forecasts of $X$.  If the transformation $T(\cdot)$  considered  is monotonic, the forecast quantiles of the transformed data will, when back-transformed, result in the correct forecast quantiles in terms of the original data. As a result finding prediction intervals in terms of the original data only requires inverting the transformation. It should be noted though, that prediction intervals that are symmetric in terms of the transformed data will not be symmetric in terms of the original data. In a similar vein, back-transformation of the forecast median of the transformed data returns the forecast median in terms of the original data. 

Back-transformation of the forecast mean of the transformed data does not yield the forecast mean of the original data, unless a linear transformation is considered. Due to the non-linearity of the transformation forecasts on the original scale of the data will be biased unless a correction is used. Bias-reduction of the forecast in the original scale has been proposed in \cite{guerrero1993time}. \cite{pankratz1987forecasts}  studied the bias of back-transforming for Box-Cox transformation. See also the work in \cite{granger1976forecasting} for forecasting with transformations. Most of the work is focused on the bias corrected mean forecasting and not in deriving the forecasting distribution of the transformed series. 

For the multivariate case see the paper \cite{arino2000forecasting} for a log-transformed VAR model with a bias correction. A simple expression for the optimal forecast under normality assumptions is derived. However, despite its theoretical advantages, the optimal forecast is shown to be inferior to the naïve forecast if specification and estimation uncertainty are taken into account \citep{baardsen2011forecasting}.

\section{An illustration}\label{the-2-dimensional-case}

We illustrate the suggested method to a 3-dimensional model of wind strength in three locations of Bergen, Norway. We have selected two locations close to the city center and one location close to the airport. The marginal distributions for the series are assumed to be Weibull with shape and scale parameters $\alpha_i$ and $\lambda_i$, respectively, $i=1,2,3$. We will here describe this model in some detail. The marginal probability density functions (PDFs) are then
\begin{equation}
f_k(x;\alpha_k,\lambda_k)=
\left\{
\begin{aligned}
& \frac{\alpha_k}{\lambda_k}\left(\frac{x}{\lambda_k}\right)^{\alpha_k-1}
e^{-\left(\frac{x}{\lambda_k}\right)^{\alpha_k}} \text{ when } x \ge 0\\
& 0 \text{ otherwise.}
\end{aligned}
\right.
\end{equation}
The CDFs can be explicitly written out as
\begin{equation}
F_k(x;\alpha_k,\lambda_k)=
\left\{
\begin{aligned}
& 1-e^{-\left(\frac{x}{\lambda_k}\right)^{\alpha_k}}, \text{ when } x \ge 0\\
& 0 \text{ otherwise.}
\end{aligned}
\right.
\end{equation}
The log-likelihood in equation \eqref{eq:loglik} can, for this particular model, be written out close to explicitly (with the exception that it involves the inverse of the standard normal CDF). The inverse from equation \eqref{eq:inverse} can be computed by
\begin{equation}
    g^{-1}(x) = \Phi^{-1}\left(1-e^{-\left(\frac{x}{\lambda}\right)^\alpha}\right)
\end{equation}
and the components to calculate \eqref{eq:jacobian} are
\begin{equation}
\frac{\partial g^{-1}(x)}{\partial x} = \frac{1}{\phi\left(\Phi^{-1}\left(1-e^{-\left(\frac{x}{\lambda}\right)^\alpha}\right)\right)}\frac{\alpha}{\lambda}\left(\frac{x}{\lambda}\right)^{\alpha-1}e^{-\left(\frac{x}{\lambda}\right)^\alpha}
\end{equation}
where the fact that
\begin{equation}
    \frac{\partial \Phi^{-1}(z)}{\partial z} = \frac{1}{\phi\left(\Phi^{-1}(z)\right)}
\end{equation}
has been used. Here $\phi(\cdot)$ denotes the PDF of the standard normal distribution. 
We use the \texttt{R}-package \texttt{TMB}, see \cite{kristensen2015tmb}, to maximize the log-likelihood. \texttt{TMB} uses automatic differentiation to find the gradients. This is extremely helpful in this illustration, and in estimating the VARTA-model in general because of the model structure with convoluted functions.

The variables are the highest daily wind strengths from the locations Florida (2 km south of the city center of Bergen), Flesland (the airport, 18km south of the city center) and Skredderdalen (2 km north-west of the city center). The names of the variables are $x_{1t}$ (Florida), $x_{2t}$ (Flesland) and $x_{3t}$ (Skredderdalen).  As we may see from  Figure \ref{fig:acf_x}, we can capture a large part of the structure by fitting  a VAR(1)-model for to $\boldsymbol{x}_t=(x_{1t},x_{2t},x_{3t})'$. The estimated model is shown in Table \ref{tab:fitted_model}. 

\begin{figure}  
    \centering
    \includegraphics[width=\textwidth,height=0.5\textheight]{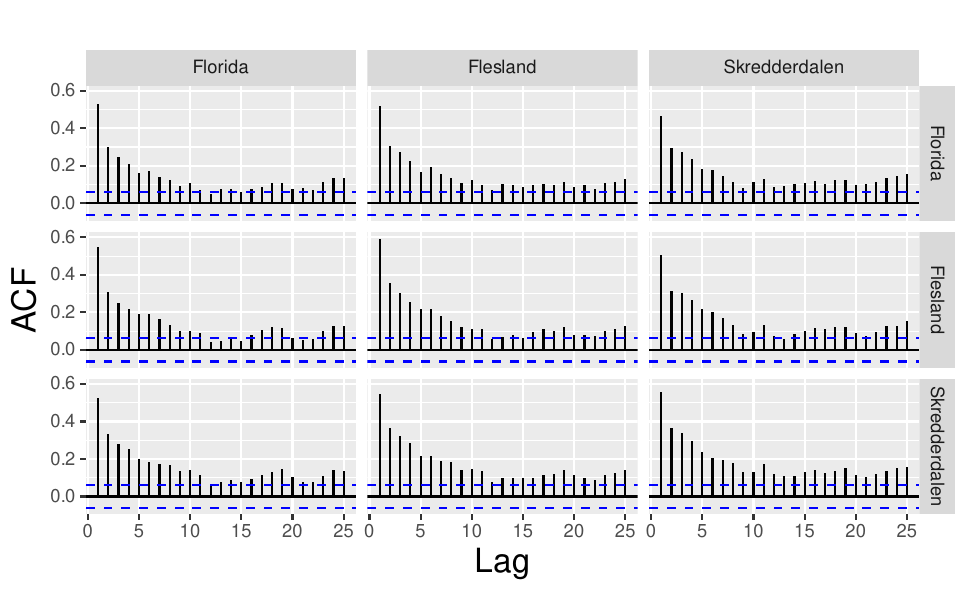}
    \caption{Auto- and cross-correlations of $\boldsymbol{x}_t$}
    \label{fig:acf_x}
\end{figure}

\begin{table}[H]
\centering
\begin{tabular}{lrrr}
  \hline
 & Parameter & st.err & t-value \\ 
  \hline
 \multicolumn{4}{c}{ Multivariate relationships}\\
  \hline
  $A_{11}$ & 0.314 & 0.080 & 3.914 \\ 
  $A_{21}$ & 0.064 & 0.075 & 0.845 \\ 
  $A_{31}$ & 0.051 & 0.076 & 0.663 \\ 
  $A_{12}$ & 0.418 & 0.080 & 5.218 \\ 
  $A_{22}$ & 0.668 & 0.075 & 8.929 \\ 
  $A_{32}$ & 0.413 & 0.075 & 5.476 \\ 
  $A_{13}$ & 0.114 & 0.064 & 1.767 \\ 
  $A_{23}$ & 0.138 & 0.060 & 2.279 \\ 
  $A_{33}$ & 0.400 & 0.062 & 6.440 \\ 
  \hline
  $\rho_{12}$ & 0.960 & 0.007 & 146.747 \\ 
  $\rho_{13}$ & 0.939 & 0.010 & 94.313 \\ 
  $\rho_{23}$ & 0.934 & 0.011 & 86.436 \\ 
  \hline
\multicolumn{4}{c}{  Parameters in the Weibull distribution}\\
  \hline
  $\alpha_1$ & 1.568 & 1.140 & 1.375 \\ 
  $\alpha_2$ & 1.469 & 1.140 & 1.289 \\ 
  $\alpha_3$ & 1.645 & 1.147 & 1.434 \\ 
  $\lambda_1$ & 12.522 & 2.600 & 4.816 \\ 
  $\lambda_2$ & 12.440 & 2.832 & 4.393 \\ 
  $\lambda_3$ & 9.644 & 2.031 & 4.750 \\ 
   \hline
\end{tabular}
\caption{An estimated  VARTA model for 3 wind speed locations in Bergen, Norway, using Weibull marginal distributions.}
\label{tab:fitted_model}
\end{table}

The results show that the wind speed on the most remote location, Flesland ($x_{2t}$), has a statistically significant effect on the wind speeds of the other two locations for the next day, see the large t-values corresponding to the coefficients $A_{12}$ and $A_{23}$. The wind speed at Florida ($x_{1t}$) does not seem to Granger-cause the wind speed on any of the other two locations, see the t-values corresponding to $A_{21}$ and $A_{31}$ while the wind speed at Skredderdalen ($x_{3t}$) Granger-cause the wind speeds at Flesland, but not Florida. 

While the results described above would also have been possible to obtain with a standard VAR(1)-model, a crucial difference is highlighted when investigating forecast distributions. An example of this, is given in Figure \ref{fig:fore_dist} where the forecast distributions for wind speed, made July 7, 2021, for the coming 9 days are shown. The distributions are obtained by the simulation based method described in Section \ref{forecasting} and we have used $M=1000$ replications. Figure \ref{fig:fore_dist} clearly shows that the forecast distributions are not Gaussian. A standard VAR-model with Gaussian errors would produce forecasts that follow Gaussian distributions, which is clearly not the case for the VARTA(1) model with Weibull-marginals.

\begin{figure}[H]
\includegraphics[width=\textwidth,height=0.5\textheight]{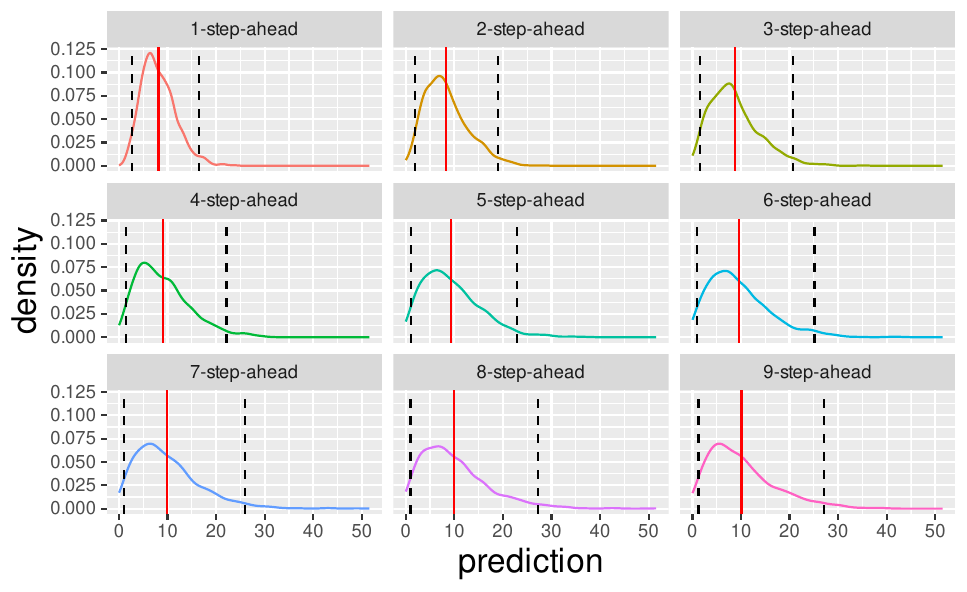}
\caption{One- to nine-step-ahead forecast distributions for wind speed on a locations near the city in Bergen, Norway. The solid vertical line represents the point prediction, i.e., the expected value of the forecast distribution. The vertical dotted lines represent a 95\% forecast intervals.}
\label{fig:fore_dist}
\end{figure}

\section{Simulations}
\label{simul}
We provide some simulation evidence to show the behavior of the proposed model. We present two experiments one in 3 dimensions  with sufficient detail and one in higher dimensions to illustrate the potential for high dimensional time series but also to see the finite sample behavior of the estimators. 
\subsection{3-dimensional data}
Trivariate series with Weibull marginals were simulated as follows.
Initially standard vector autoregressive model was simulated
with 
\[
{\bf A} = \left[ 
\begin{array}{ccc}
0.7 &0.2&0.1 \\
0.3&0.5&0.2 \\
0.1&0.7&-0.2
\end{array}
\right]
\]
and 
\[
{\bf \Sigma} = 
\left[ 
\begin{array}{ccc}
1 &\rho_{12}&\rho_{13} \\
&1&\rho_{23} \\
&&1
\end{array}
\right] = \left[ 
\begin{array}{ccc}
1 &0.5&0.3 \\
&1&0.7 \\
&&1
\end{array}
\right].
\]
Denote the simulated series as $Z_{it}$, $i=1,2,3$ and $t=1,\ldots,n$.
Then the  series were transformed back to three
Weibull series by applying the 
inverse transformation, namely
\[
X_{it}= F^{-1} (\Phi(Z_{it});\alpha,\beta)
\]
where $F^{-1}(\cdot;\alpha,\beta)$ 
is the quantile function of a Weibull distribution
with parameters $\alpha$ and $\beta$
and $\Phi(\cdot)$ is the CDF of the standard Gaussian distribution.
We have used
$\alpha_1=2, \alpha_2=2, \alpha_3=3$ and $\beta_1=3, \beta_2=5, \beta_3=1$. 
Also we have used for the length of the series $n=200,500,1000,5000$. 

For each data set generated we estimated the  parameters of interest applying ML approach as described in Section \ref{inference}.

To facilitate the calculations we made use of the \texttt{TMB} package in \texttt{R} \citep{kristensen2015tmb}.
The asymptotic standard errors, computed by \texttt{TMB}, was used to compute 95\% confidence intervals for each replication and parameter.

\begin{center}
\begin{figure}
\includegraphics[scale=0.54]{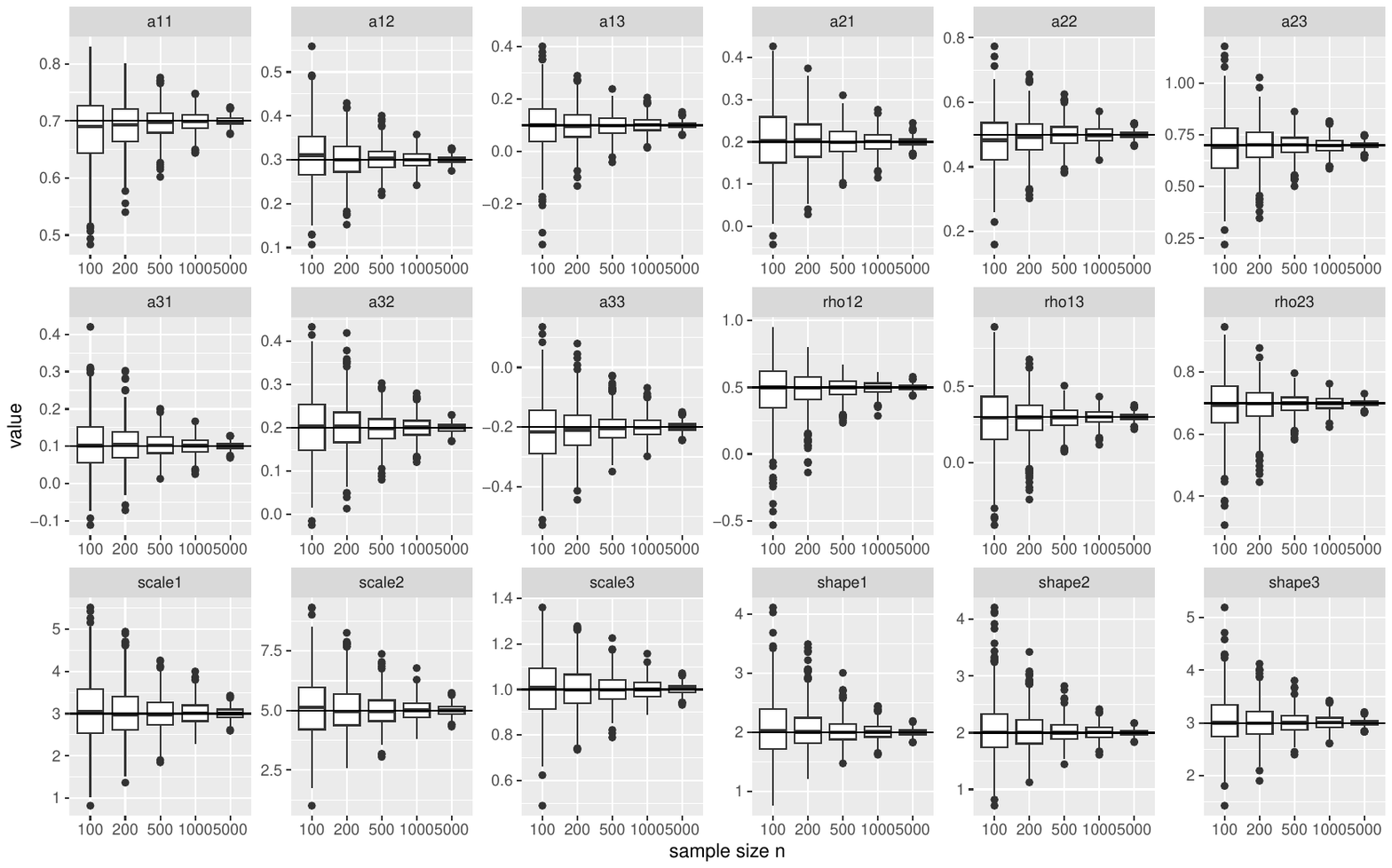}
\caption{\label{simfig} Boxplot of the estimated parameters based on 1000 replications in the 3-dimensional case. }
\end{figure}
\end{center}

Figure \ref{simfig} depicts the results from the simulations for each of the parameters involves. The horizontal line indicates the true underlying value used for the simulations. The behavior is as expected. We see that increasing the sample size the bias is negligible while the standard errors decrease. 
Overall the parameters are estimated very well by ML.

\begin{table}[ht]
\centering
\begin{tabular}{rrrrr}
  \hline
 & $n=200$ & $n=500$ & $n=1000$ & $n=5000$ \\ 
  \hline
   \multicolumn{5}{c}{ Multivariate relationships}\\
  \hline
$A_{11}$ & 0.959 & 0.939 & 0.947 & 0.949 \\ 
$A_{21}$   & 0.960 & 0.958 & 0.949 & 0.956 \\ 
$A_{31}$ & 0.951 & 0.963 & 0.935 & 0.949 \\ 
$A_{12}$   & 0.942 & 0.936 & 0.947 & 0.959 \\ 
$A_{22}$   & 0.961 & 0.951 & 0.946 & 0.951 \\ 
$A_{32}$   & 0.941 & 0.945 & 0.950 & 0.952 \\ 
$A_{13}$   & 0.957 & 0.960 & 0.953 & 0.961 \\ 
$A_{23}$   & 0.951 & 0.935 & 0.947 & 0.955 \\ 
$A_{33}$   & 0.945 & 0.949 & 0.938 & 0.940 \\ 
\hline
 $\rho_{12}$ & 0.923 & 0.944 & 0.954 & 0.958 \\ 
 $\rho_{13}$ & 0.927 & 0.953 & 0.948 & 0.954 \\ 
 $\rho_{23}$ & 0.938 & 0.951 & 0.952 & 0.940 \\ 
\hline
\multicolumn{5}{c}{  Parameters in the Weibull distribution}\\
  \hline
$\alpha_1$ & 0.928 & 0.934 & 0.951 & 0.942 \\ 
$\alpha_2$ & 0.924 & 0.935 & 0.955 & 0.946 \\ 
$\alpha_3$ & 0.937 & 0.944 & 0.946 & 0.946 \\ 
$\lambda_1$ & 0.919 & 0.945 & 0.951 & 0.947 \\ 
$\lambda_2$& 0.927 & 0.939 & 0.955 & 0.945 \\ 
$\lambda_3$& 0.923 & 0.940 & 0.954 & 0.943 \\ 
 \hline
Average & 0.940 & 0.946 & 0.949 & 0.950 \\ 
\hline
\end{tabular}
\caption{\label{coverage} Empirical coverage for 95\%
confidence intervals based on the asymptotic results.
Data were generated from a trivariate model with Weibull marginals }
\end{table}

Initial values were selected as follows: for the VAR parameters we run a
false VAR model to capture the structure. For the parameters of the Weibull distributions we fitted marginal Weibull models.  Finally, for the correlation parameters we estimated them from the series ignoring the temporal nature of the data. Monitoring the convergence of the optimizer we never found a non-convergent case. 

Table \ref{coverage} reports the empirical coverage of the 95\% confidence intervals for all model parameters, computed under the assumption of asymptotic normality. Empirical coverage refers to the proportion of replications in which the true parameter value lies within the estimated confidence interval.
The results are based on 1000 simulation replications. The reported coverage rates are consistently close to the nominal level of 95\%, suggesting that the asymptotic approximation performs well even for relatively small sample sizes (e.g., $n=100$).

The final row presents the average empirical coverage across all parameters, demonstrating that, as the sample size increases, the coverage rates converge toward the nominal 95\% level.

\subsection{Higher dimension}
We have also run some simulations with $d=6$.
Here matrix $\bf A$ has elements $A_{ij}=0.1$ for $i,j=1,\ldots,6$ and
\begin{equation}
{\bf \Sigma} = \left[ 
\begin{array}{cccccc}
  1 & 0.1 & 0.1 & 0.1 & 0.1 & 0.1 \\ 
    0.1 & 1 & 0.4 & 0.1 & 0.1 & 0.4 \\ 
    0.1 & 0.4 & 1 & 0.4 & 0.1 & 0.1 \\ 
    0.4 & 0.1 & 0.1 & 1 & 0.1 & 0.1 \\ 
    0.1 & 0.1 & 0.1 & 0.1 & 1 & 0.1 \\ 
    0.1 & 0.1 & 0.4 & 0.1 & 0.1 & 1 \\ 
 \end{array}
\right]
\end{equation}
while shape and scale parameters for the Weibull distributions were
$\alpha= (2,2,3,2,2,3)$ and 
$\beta=(3,5,1,2,4,6)$. Overall we have to estimate 63 parameters.

\begin{table}[ht]
\centering
\begin{tabular}{ccccc}
  \hline
    &\multicolumn{3}{c}{ Parameter set} \\
 sample size& ${\bf A}$  &  Weib par. & $\rho$ & All \\ 
  \hline
$n=100$ & 0.937 & 0.942 & 0.939 & 0.938 \\ 
 $n=200$  & 0.943 & 0.937 & 0.945 & 0.942 \\ 
$n=500$ & 0.945 & 0.946 & 0.953 & 0.947 \\ 
$n=1000$ & 0.949 & 0.940 & 0.952 & 0.948 \\ 
$n=5000$ & 0.952 & 0.951 & 0.952 & 0.952 \\ 
   \hline
\end{tabular}
\caption{\label{empir} Empirical coverage for 95\% confidence intervals for the parameters, for the  case with $d=6$. We have organized the output with respect the three different groups of parameters. The first one relates to the matrix $\bf A$, the second one to the Weibull parameters and the third one the parameters from the correlation matrix. The last column is the mean over all parameters. We see that the empirical coverage is very close to the nominal one. }
\end{table}

Table \ref{empir} presents the empirical coverage of the 95\% confidence intervals for the model parameters. The results are organized into three groups. The first group corresponds to the parameters of the matrix $\bf A$ (36 parameters),
the second group includes the Weibull parameters — both scale and shape — totaling 12 parameters and the  third group consists of the parameters from the correlation matrix (15 parameters).
The final column reports the average empirical coverage across all parameters. In all cases, the empirical coverage is very close to the nominal 95\% level, indicating good finite-sample performance of the asymptotic confidence intervals.


Figure \ref{MSEs} shows the root mean squared error (RMSE) for the parameters as a function of the sample size. Again we have organize the plot in three groups of parameters.
Let $\theta_j$, $j=1,\ldots,J$ be the parameters of interest.
We report the RMSE as
\[
\text{RMSE} =
\sqrt{ \frac{1}{J} \sum\limits_{j=1}^{J} \left( \frac{1}{R} 
\sum_{r=1}^R \left( \hat{\theta}_j^{(r)} - \theta_j \right)^2 \right),
}
\]
where $R$ is the number of replications, in our case $R=1000$,
and 
$\hat{\theta}_j^{(r)}  $ is the estimated value at the $r$th replication.
The behavior as expected shows that the estimators are satisfactory.

\begin{center}
\begin{figure}
\includegraphics[scale=0.49]{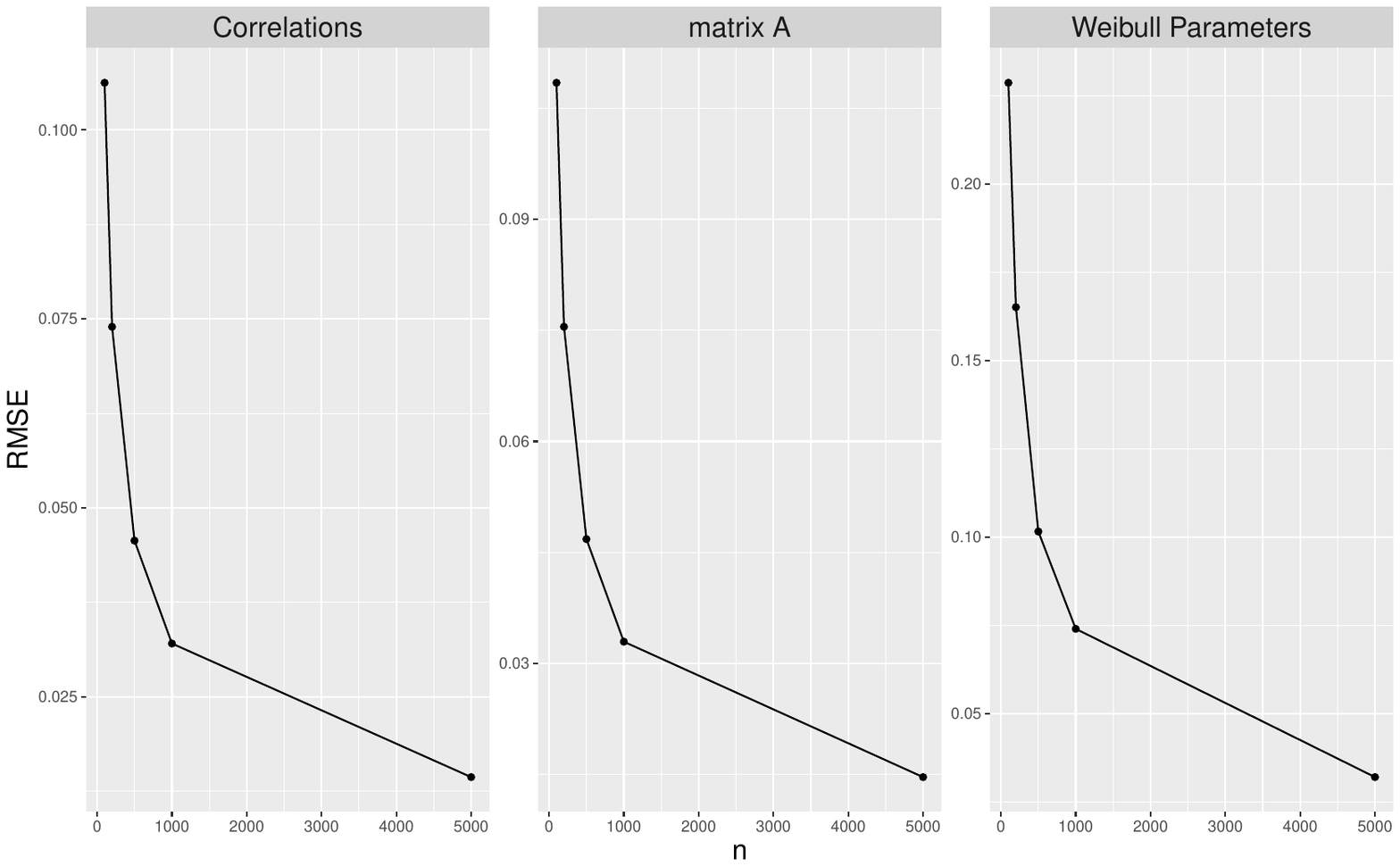}
\caption{\label{MSEs} Root MSE of the estimated parameters based on 1000 replications in the 3-dimensional case. }
\end{figure}
\end{center}

\section{Concluding Remarks }
\label{conclude}

In this paper, we considered inferential methodology for time series of non-Gaussian data by employing a transformation combined with a vector autoregressive (VAR) model. 
A key ingredient of our approach is its flexibility: it allows for different marginal distributions to be incorporated in a unified framework. Moreover, since our method provides access to the log-likelihood function, it enables the use of standard likelihood-based tools such as model comparison and order selection. Also, note that since the full forecasting distribution can be computed, one can use the model for any kind forecasts, e.g. a point forecast or an interval forecast.

The proposed framework can be extended to other data types, including count time series. For instance, in \cite{jia2021latent}, the idea of assuming an underlying latent Gaussian process was explored in the univariate setting. A similar concept was also used in \cite{masarotto2012gaussian}. Our approach generalizes this idea to higher dimensions in a natural way.

In the case of count time series, the transformation is not unique, and identification of the transformation may require additional restrictions. Nevertheless, the methodology can still be applied effectively, with only minor modifications. This opens the possibility of handling mixed-type data by incorporating copula models to link different series, followed by transformation using our proposed method.

Model misspecification is a well-known issue in time series analysis, particularly when fitting an autoregressive model without knowledge of the true order of the underlying process. In such cases, the model is likely to be misspecified, leading to inconsistent parameter estimates (see, e.g., \cite{bhansali1981effects}). This issue is relevant in our framework as well, since the true order is typically unknown in practical applications. Investigating the effects of misspecification within our approach is therefore an important direction for future research. Importantly, since our method allows for more flexible assumptions on the marginal distributions, it may help mitigate biases arising from inappropriate normality assumptions.

\bibliographystyle{apalike}
\bibliography{references}

\end{document}